# VIBRATIONAL MODE ANALYSIS OF VOID INDUCED CORONENE AS A POSSIBLE CARRIER OF THE ASTRONOMICAL AROMATIC INFRARED BANDS

NORIO OTA

[1] Graduate School of Pure and Applied Sciences, University of Tsukuba,
1-1-1 Tenoudai Tsukuba-city 305-8571, Japan; n-otajitaku@nifty.com

Void induced di-cation coronene $C_{23}H_{12}^{++}$ is a possible carrier of the astronomically observed polycyclic aromatic hydrocarbon (PAH). Based on density functional theory, multiple spin state analysis was done for neutral void coronene $C_{23}H_{12}$. Singlet spin state was most stable (lowest total energy). By the Jahn-Teller effect, there occurs serious molecular deformation. Point group D6h of pure coronene transformed to C2 symmetry having carbon two pentagons. Advanced singlet stable molecules were di-cation $C_{23}H_{12}^{++}$ and di-anion $C_{23}H_{12}^{--}$. Molecular configuration was almost similar with neutral $C_{23}H_{12}$. However, electric dipole moment of these two charged molecules show reversed direction with 1.19 and 2.63 Debey. Calculated infrared spectrum of $C_{23}H_{12}^{++}$ show a very likeness to observed one of two astronomical sources of HD44179 and NGC7027. Harmonic vibrational mode analysis was done for $C_{23}H_{12}^{++}$. At 3.2 µm, C-H stretching at pentagons was featured. From 6.4 to 8.7µm, C-C stretching mode was observed. In-plane-bending of C-H was in a range of 7.6-9.2µm. Both C-H out-of plane bending and C-C stretching were accompanied from 11.4 to 14.3µm. Astronomically observed emission peaks of 3.3, 6.2, 7.6, 7.8, 8.6, 11.2, 12.7, 13.5 and 14.3µm were compared well with calculated peaks of 3.2, 6.5, 7.6, 7.8, 8.6, 11.4, 12.9, 13.5, and 14.4µm.

## 1, INTRODUCTION

Interstellar dust show mid-infrared emission from 3 to 20µm. Discrete emission features at 3.3, 6.2, 7.6, 7.8, 8.6, 11.2, and 12.7µm are ubiquitous peaks observed at many astronomical objects (Ricca et al. 2012; Geballeet al. 1989; Verstraete et al. 1996; Moutou et al. 1999; Meeus et al. 2001; Peeters et al. 2002; Regan et al.2004; Engelbracht et al. 2006; Armus et al. 2007; Smith et al. 2007; Sellgren et al. 2007). Current understanding is that these astronomical spectra come from the vibrational modes of polycyclic aromatic hydrocarbon (PAH) molecules. Concerning PAH spectra, there are many experimental (Szczepanski & Vala 1993a; Schlemmer et al. 1994; Moutou et al. 1996; Cook et al. 1998; Piest et al. 1999; Hudgins& Allamandola 1999a, 1999b; Oomens et al. 2001, 2003, 2011;Kim et al. 2001) and density functional theory (DFT) based theoretical analysis (de Frees et al. 1993; Langhoff 1996;Mallocci et al. 2007; Pathak & Rastogi 2007; Bauschlicher et al.2008, 2009; Ricca et al. 2010, 2011b, 2012).

The current central concept to understand the observed astronomical spectra is the decomposition method from the data base of many PAHs experimental and theoretical analysis (Boersma et al. 2013, 2014). Recently, Tielens (Tielens 2013) discussed void induced graphene like PAH's, which may be candidates of emission source. In a previous paper (Ota 2014b), void induced coronene $C_{23}H_{12}^{++}$ show a very likeness to well observed one by using DFT calculation method. The aim of this paper is to analyze molecule harmonic vibration mode of $C_{23}H_{12}^{++}$. Also, multiple spin state of void coronene was studied to search serious Jahn-Teller effect resulting characteristic molecular configuration and fairly large electronic dipole.

## 2, CALCULATION METHOD

We have to obtain total energy, optimized atom configuration, and infrared vibrational mode frequency and strength depend on a given initial atomic configuration, charge and spin state Sz. Density functional theory (DFT) with unrestricted B3LYP functional (Becke 1993) was applied utilizing Gaussian09 package (Frisch et al. 2009, 1984) employing an atomic orbital 6-31G basis set. The first step calculation is to obtain the self-consistent energy, optimized atomic configuration and spin density. T R Required convergence on the root mean square density matrix was less than $10^{-8}$ within 128 cycles. Based on such optimized results, atomic vibrational frequency and strength was calculated. Vibration strength is obtained as molar absorption coefficient ε (km/mol). Comparing DFT harmonic wavenumber $N_{DFT}(cm^{-1})$ with experimental data, a single scale factor 0.958 was used (Ricca et al. 2012). Observed spectra are astronomical PAHs emission. As noted details by Ricca et al. (Ricca et al. 2012), we should consider photon absorbed emission and apply red shift by 15 cm$^{-1}$.

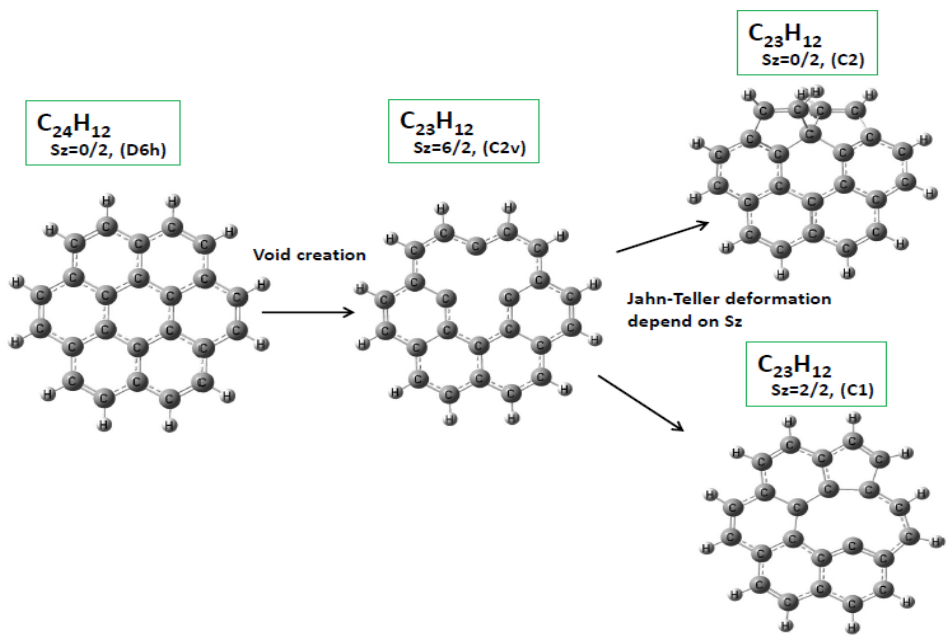

**Figure 1.** Void induced coronene family. Pure coronene has D6h point group symmetry. Optimized molecular configuration depends on spin state Sz by the Jahn-Teller effect. Singlet spin state (Sz=0/2) of $C_{23}H_{12}$ show group symmetry C2 with carbon two pentagons, whereas triplet (Sz=2/2) C1 with one pentagon.

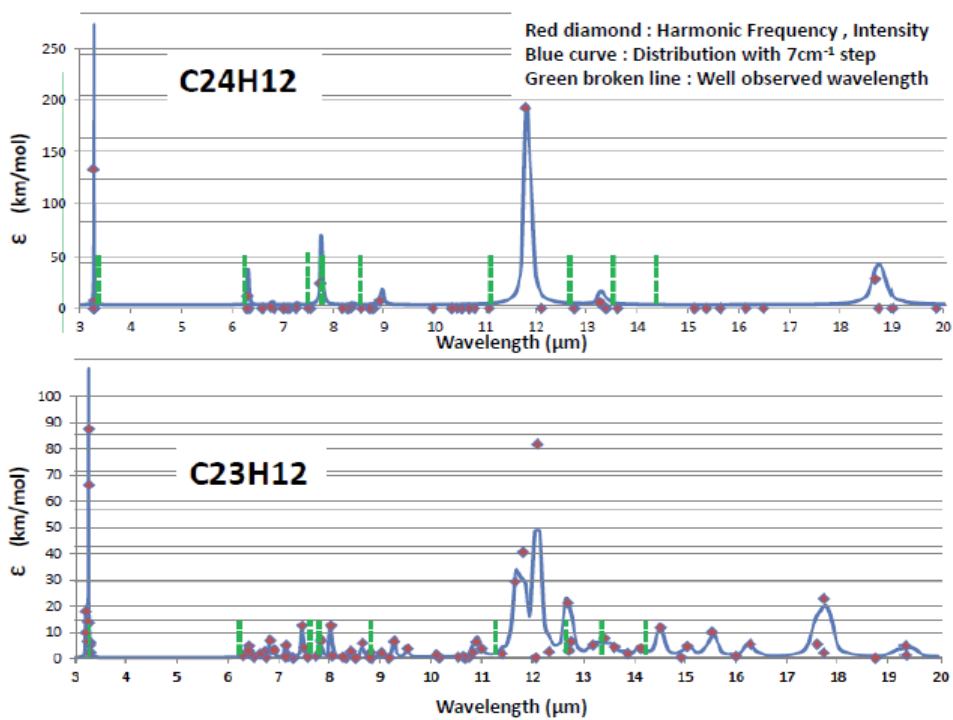

**Figure 2.** Calculated and corrected infrared spectra of $C_{24}H_{12}$ and singlet spin state $C_{23}H_{12}$. Red diamond show harmonic wavelength (micrometer) and intensity epsiron (km/mol). Blue curve is its distribution. Green broken lines are well observed wavelength in interstellar dust.

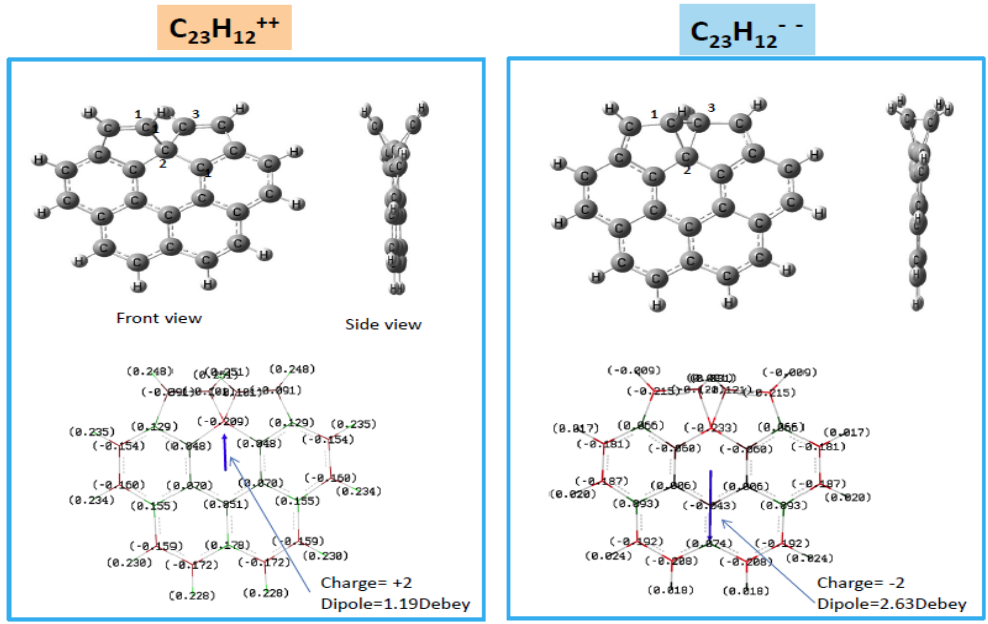

**Figure 3.** Comparing di-cation C23H12++ with di-anion C23H12--, molecular configuration is almost similar with carbon two pentagons. However, electric dipole moment (blue arrow) show reversed direction with 1.19 and 2.63Debey.

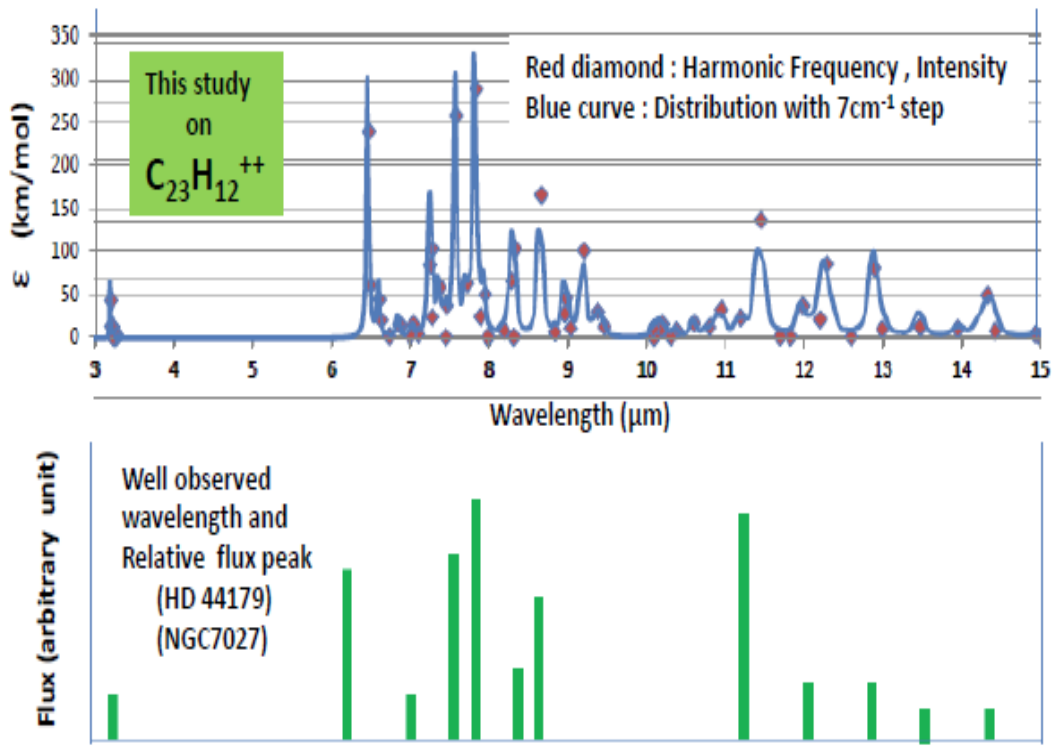

**Figure 4.** Calculated harmonic wavelength and intensity of $C_{23}H_{12}^{++}$ is shown by red diamond and their distribution curve by a blue curve based on every 7 cm$^{-1}$ accumulation step. Green bar is observed flux peak position and height relatively estimated through two astronomical sources of HD44179 and NGC7027 (modified original spectra; Boersma et al. 2009).

Corrected wave number N is obtained simply by,

$N(cm^{-1}) = N_{DFT}(cm^{-1}) \times 0.958 - 15 \ (cm^{-1})$

Also, wavelength λ is,

$\lambda(\mu m) = 10000/N(cm^{-1})$

In order to compare with reported calculations on pure coronene spectrum, Appendix 1 is inserted in this paper, where wavelength is corrected one.

3, MULTIPLE SPIN STATE AND JAHN-TELLER DISTORTION

Void coronene molecules were illustrated in Figure 1. Pure coronene $C_{24}H_{12}$ is non-magnetic (Sz=0/2) having D6h point group symmetry.

Creation of carbon single void was supposed as shown in middle part of Figure 1 as $C_{23}H_{12}$. There are six unpaired electrons, which means multiple spin state capability of Sz=6/2, 4/2, 2/2 and 0/2. We should find which one is the most stable (lowest total energy).

Table 1 was a result comparing energy difference between spin state Sz. Most stable one was singlet state (Sz=0/2). Next is triplet (Sz=2/2) with 70 kcal/mol higher than singlet. Also, we should compute molecular configuration depend on spin state, that is, such a void brings quantum mechanical distortion by the Jahn-Teller effect (Ota; 2011, 2014a). It is amazing that there cause bond-bond reconstruction. In case of singlet spin state (right upper part in Figure 1), we can see carbon two pentagons connected with five hexagons with a point group symmetry of C2. Whereas, in triplet case (lower part), one pentagon was observed with C1 point group.

**Table 1**

Multiple spin state analysis of neutral void coronene $C_{23}H_{12}$. By a creation of single carbon void, there brings four spin states. Most stable (lowest total energy) one was Sz=0/2 (singlet). Increasing spin parameter Sz, total energy increases and became unstable

| Input Sz | Calculated S(S+1) | Energy difference (kcal/mol) | Molecule Point group |
|---|---|---|---|
| 0/2 | 0.00 | 0 | C2 |
| 2/2 | 2.05 | 70 | C1 |
| 4/2 | 6.13 | 130 | C1 |
| 6/2 | 12.04 | 208 | C2v |

4. Di-CATION $C_{23}H_{12}^{++}$ AND Di-ANION $C_{23}H_{12}^{--}$

Unfortunately, infrared spectrum of singlet state neutral void coronene $C_{23}H_{12}$ was far from observed one as shown in Figure 2. We should find other candidate. One idea is spin canceling mechanism considering molecular orbits. Advanced singlet spin state molecules will be di-cation $C_{23}H_{12}^{++}$ and di-anion $C_{23}H_{12}^{--}$. Because, by pulling out two electrons from HOMO level of singlet $C_{23}H_{12}$ we can realize singlet state to be $C_{23}H_{12}^{++}$. Whereas, by adding two electrons to LUMO level we can realize again singlet state as $C_{23}H_{12}^{--}$.

Optimized configuration was illustrated in Figure 3, which show similar configuration in both cases, but somewhat detailed discrepancy on two pentagons connected angle $C_1$-$C_2$-$C_3$ (see Figure 3 attached number beside carbons). In case of $C_{23}H_{12}^{++}$, it was 95°, while $C_{23}H_{12}^{--}$ 58° which is smaller enough to make an extra sigma bonding between $C_1$ and $C_3$. Important issue is a charge distribution and electric dipole moment. Lower part of Figure 3 show charge distribution by a small letter. In case of $C_{23}H_{12}^{++}$, hydrogen has plus charge of 0.22e to 0.25e, while $C_{23}H_{12}^{--}$ small value of 0.01e to 0.02e. Electric dipole reflects such distribution. Dipole of $C_{23}H_{12}^{++}$ was 1.19 Debey oriented to upward in a molecule-plane, whereas that of $C_{23}H_{12}^{--}$ was 2.63 Debey downward vector.

5, MID-IR SPECTRUM OF $C_{23}H_{12}^{++}$

In Figure 4, Calculated harmonic wavelength and intensity of $C_{23}H_{12}^{++}$ is shown by red diamond and their distribution curve by blue accumulating every 7 cm$^{-1}$ step. Green bar is observed flux peak position and height relatively estimated through two astronomical sources of HD44179 and NGC7027. Original full observed data was opened by Boersma et al. (Boersma et al 2009). We can see good coincidence between two. Of course, it should be noted that observation is emission flux, while this calculation show absorption. We need advanced calculation on emission using harmonic vibration results appeared in Appendix 2. Detailed data were shown in Table 2. Observed peaks are 3.3, 6.2, 7.6, 7.8, 8.6, 11.2, 12.7, 13.5 and 14.3μm, whereas calculated peaks 3.2, 6.5, 7.3, 7.6, 7.8, 8.3, 8.7, 9.0, 9.2, 11.4, 12.9, 13.5, and 14.3μm respectively. We cannot identify 9.0 and 9.2μm calculated peaks in observed one. Also, we should notice a large discrepancy at 11.2μm band. Calculated relative intensity is almost one third of observed one.

More inspection and study is necessary to have a complete set of molecules. Anyway, this study could present almost good coincidence by a single molecule $C_{23}H_{12}^{++}$.

Concerning di-anion $C_{23}H_{12}^{--}$, unfortunately, calculated spectrum (Ota 2014b) was very different with observed one.

Table 2

Vibrational mode analysis of $C_{23}H_{12}^{++}$. Large intensity modes were selected among full 99 modes noted in appendix 2. Calculated and corrected (scale factor and red shift) wavelength was compared with well observed one in interstellar dust.

| Observation | Calculation for $C_{23}H_{12}^{++}$ | | | notes |
|---|---|---|---|---|
| Well observed wavelength (μm) | wavelength (μm) | Vibration mode | Harmonic number (Appendix2) | |
| 3.3 | 3.19 | C-H stretching at pentagons | 98 | |
| 6.2 | 6.46 | C-C stretching at hexagons | 87 | |
| | 7.27 | C-C stretching at all carbons | 75 | Not identified |
| 7.6 | 7.57 | C-H in-plane bending, C-C stretching | 71 | |
| 7.8 | 7.82 | C-H in-plane bending, C-C stretching | 69 | |
| | 7.93 | C-H in-plane bending, C-C stretching | 67 | |
| | 8.32 | C-H in-plane bending, C-C stretching | 62 | Uncertain Observation |
| 8.6 | 8.65 | C-H in-plane bending, C-C stretching | 61 | |
| | 8.97 | C-H in-plane bending | 58 | Not identified |
| | 9.19 | C-H in-plane bending | 56 | Not identified |
| 11.2 | 11.44 | C-H out-of-plane bending, C-C stretching | 43 | |
| | 12.28 | C-H out-of-plane bending, C-C stretching | 38 | Uncertain |
| 12.7 | 12.88 | C-H out-of-plane bending, C-C stretching | 36 | |
| 13.5 | 13.47 | C-H out-of-plane bending at pentagons | 33 | |
| 14.3 | 14.32 | C-H out-of plane bending at pentagons, C-C stretching | 31 | |
| 17.1 | 17.09 | Carbon out-of-planebending | 25 | |
| | 22.14 | Hexagon in-plane twisting | 17 | |
| | 31.01 | Hexagon in-plane twisting | 11 | |
| | 56.37 | Molecule out-of-plane bending | 5 | |

## Appendix 1. $C_{24}H_{12}$ fundamental vibrations

| C24H12 | D6h point group symmetry | | | | | | | | | | |
|---|---|---|---|---|---|---|---|---|---|---|---|
| Harmonic number | symmetry | Wavelength (μm) | IR Intensity (km/mol) | | | | | | | | |
| 1 | E2U | 136.699517 | 0 | 31 | E2G | 15.1289339 | 0 | 67 | A2G | 8.18510875 | 0 |
| 2 | E2U | 136.699517 | 0 | 32 | E2U | 13.6153375 | 0 | 68 | E1U | 7.73764214 | 24.3467 |
| 3 | A2U | 91.1560427 | 5.0256 | 33 | E2U | 13.6153375 | 0 | 69 | E1U | 7.73764214 | 24.3464 |
| 4 | B1G | 67.8436339 | 0 | 34 | B1G | 13.3998949 | 0 | 70 | B2U | 7.55650469 | 0 |
| 5 | B2G | 47.5113269 | 0 | 35 | B2G | 13.3656672 | 0 | 71 | A1G | 7.49973040 | 0 |
| 6 | E1G | 36.114981 | 0 | 36 | E1U | 13.2604932 | 5.6576 | 72 | E2G | 7.26601933 | 0 |
| 7 | E1G | 36.114981 | 0 | 37 | E1U | 13.2604932 | 5.6573 | 73 | E2G | 7.26601933 | 0 |
| 8 | E2U | 35.3787579 | 0 | 38 | E1U | 12.7646716 | 0.0659 | 74 | E1U | 7.26303698 | 0.9008 |
| 9 | E2U | 35.3787579 | 0 | 39 | E1U | 12.7646716 | 0.066 | 75 | E1U | 7.26303698 | 0.901 |
| 10 | E2G | 28.8572000 | 0 | 40 | E2U | 12.7436726 | 0 | 76 | B1U | 7.13856334 | 0 |
| 11 | E2G | 28.8572000 | 0 | 41 | E2U | 12.7436726 | 0 | 77 | E2G | 7.09505964 | 0 |
| 12 | E1U | 27.6894153 | 3.2384 | 42 | E1G | 12.1107593 | 0 | 78 | E2G | 7.09505964 | 0 |
| 13 | E1U | 27.6894153 | 3.2386 | 43 | E1G | 12.1107593 | 0 | 79 | E2G | 7.01163618 | 0 |
| 14 | E1G | 23.1468329 | 0 | 44 | A2U | 11.7981278 | 192.3288 | 80 | E2G | 7.01163618 | 0 |
| 15 | E1G | 23.1468329 | 0 | 45 | A2G | 11.0741072 | 0 | 81 | B2U | 6.83597544 | 0 |
| 16 | A1G | 22.1075224 | 0 | 46 | A1U | 10.7907109 | 0 | 82 | E1U | 6.77066064 | 1.4942 |
| 17 | B2U | 21.8765877 | 0 | 47 | E1G | 10.6879757 | 0 | 83 | E1U | 6.77066064 | 1.4941 |
| 18 | E2G | 21.1538499 | 0 | 48 | E1G | 10.6879757 | 0 | 84 | A2G | 6.60787099 | 0 |
| 19 | E2G | 21.1538499 | 0 | 49 | E2U | 10.5393755 | 0 | 85 | B1U | 6.59224726 | 0 |
| 20 | A1U | 19.8763211 | 0 | 50 | E2U | 10.5393755 | 0 | 86 | A1G | 6.35448932 | 0 |
| 21 | E2U | 19.0190089 | 0 | 51 | B2G | 10.4517908 | 0 | 87 | E2G | 6.30083521 | 0 |
| 22 | E2U | 19.0190089 | 0 | 52 | E2G | 10.3418619 | 0 | 88 | E2G | 6.30083521 | 0 |
| 23 | B1U | 18.7696828 | 0 | 53 | E2G | 10.3418619 | 0 | 89 | E1U | 6.2992348 | 12.1109 |
| 24 | A2U | 18.6925472 | 28.2844 | 54 | A1G | 9.96529047 | 0 | 90 | E1U | 6.2992348 | 12.1117 |
| 25 | B2G | 16.4914584 | 0 | 55 | E1U | 8.92372919 | 7.4894 | 91 | A2G | 3.29392677 | 0 |
| 26 | A2G | 16.1440355 | 0 | 56 | E1U | 8.92372919 | 7.4907 | 92 | E1U | 3.29304381 | 7.5751 |
| 27 | E1G | 15.6390195 | 0 | 57 | B2U | 8.79944182 | 0 | 93 | E1U | 3.29304381 | 7.5769 |
| 28 | E1G | 15.6390195 | 0 | 58 | B1U | 8.77254423 | 0 | 94 | E2G | 3.29138422 | 0 |
| 29 | B1U | 15.3660987 | 0 | 59 | E2G | 8.71591876 | 0 | 95 | E2G | 3.29138422 | 0 |
| 30 | E2G | 15.1289339 | 0 | 60 | E2G | 8.71591876 | 0 | 96 | B1U | 3.29065013 | 0 |
| | | | | 61 | B2U | 8.55575688 | 0 | 97 | B2U | 3.27247243 | 0 |
| | | | | 62 | E1U | 8.35950295 | 0.779 | 98 | E2G | 3.27133845 | 0 |
| | | | | 63 | E1U | 8.35950295 | 0.7795 | 99 | E2G | 3.27133845 | 0 |
| | | | | 64 | A1G | 8.30114047 | 0 | 100 | E1U | 3.26932595 | 133.2718 |
| | | | | 65 | E2G | 8.27591916 | 0 | 101 | E1U | 3.26932595 | 133.2895 |
| | | | | 66 | E2G | 8.27591916 | 0 | 102 | A1G | 3.26764374 | 0 |

Symmetry, A1G:$A_{1g}$ A2G:$A_{2g}$ B1G:$B_{1g}$ B2G:$B_{2g}$ E1G:$E_{1g}$ E2G:$E_{2g}$ A1U:$A_{1u}$ A2U:$A_{2u}$ B1U:$B_{1u}$ B2U:$B_{2u}$ E1U:$E_{1u}$ E2U:$E_{2u}$

(Above wavelength are DFT calculated and corrected by scale factor 0.958 and red shift by 15cm$^{-1}$.)

## Appendix 2. $C_{23}H_{12}^{++}$ fundamental vibrations

| Harmonic number | C2 point group symmetry | wavelength (μm) | IR intensity (km/mol) |
|---|---|---|---|
| 1 | B | 128.440089 | 4.5889 |
| 2 | A | 120.958215 | 1.3381 |
| 3 | B | 101.006008 | 11.8229 |
| 4 | A | 62.1307205 | 0.0878 |
| 5 | B | 56.3712692 | 55.1832 |
| 6 | A | 45.7129869 | 8.376 |
| 7 | B | 41.5036719 | 11.4499 |
| 8 | A | 36.2784414 | 0.2552 |
| 9 | A | 34.9079549 | 0.2215 |
| 10 | B | 33.9062235 | 3.1309 |
| 11 | B | 31.0090831 | 86.9236 |
| 12 | A | 29.0964620 | 11.6713 |
| 13 | B | 25.9762776 | 0.905 |
| 14 | A | 23.7847460 | 0.003 |
| 15 | B | 23.4045730 | 2.1241 |
| 16 | A | 23.2859605 | 2.8943 |
| 17 | B | 22.1422764 | 46.4723 |
| 18 | A | 21.4372883 | 1.1949 |
| 19 | B | 21.0467167 | 3.6327 |
| 20 | A | 20.3045203 | 1.3847 |
| 21 | A | 19.4352599 | 13.3558 |
| 22 | B | 18.7795533 | 17.3062 |
| 23 | A | 18.2282356 | 0.31 |
| 24 | B | 17.8177725 | 8.1806 |
| 25 | B | 17.0906194 | 45.5637 |
| 26 | A | 16.1129599 | 1.2534 |
| 27 | A | 15.8663897 | 0.076 |
| 28 | A | 14.9911342 | 0.265 |
| 29 | B | 14.9337759 | 3.7441 |
| 30 | B | 14.4194092 | 7.1985 |
| 31 | B | 14.3209489 | 48.8426 |
| 32 | A | 13.9463428 | 9.6893 |
| 33 | B | 13.4751733 | 11.4881 |
| 34 | A | 13.4659461 | 12.0032 |
| 35 | A | 12.9848813 | 9.2307 |
| 36 | B | 12.8866774 | 80.717 |
| 37 | A | 12.5909407 | 0.2628 |
| 38 | B | 12.2823723 | 85.3515 |
| 39 | A | 12.1964302 | 20.7214 |
| 40 | B | 11.9792603 | 36.3086 |
| 41 | A | 11.8196108 | 0.7153 |
| 42 | B | 11.6879829 | 1.0484 |
| 43 | B | 11.4444167 | 136.2236 |
| 44 | B | 11.1885463 | 23.0379 |
| 45 | A | 10.9410211 | 32.1274 |
| 46 | B | 10.7894572 | 11.7865 |
| 47 | A | 10.595496 | 15.8518 |
| 48 | A | 10.375279 | 7.64 |
| 49 | B | 10.3010348 | 0.8388 |
| 50 | B | 10.196156 | 15.6457 |
| 51 | A | 10.1780281 | 8.7556 |
| 52 | B | 10.0862667 | 0.0246 |
| 53 | A | 10.0783465 | 10.0546 |
| 54 | B | 9.45675166 | 12.2548 |
| 55 | A | 9.37179293 | 30.0139 |
| 56 | B | 9.19611578 | 100.7882 |
| 57 | A | 9.02729871 | 10.4829 |
| 58 | A | 8.96936286 | 45.5088 |
| 59 | B | 8.95409808 | 27.1882 |
| 60 | A | 8.82668996 | 5.762 |
| 61 | B | 8.65430516 | 165.7545 |
| 62 | A | 8.3208072 | 103.2221 |
| 63 | A | 8.30433814 | 0.4494 |
| 64 | B | 8.27824912 | 65.8739 |
| 65 | B | 8.18188999 | 7.8531 |
| 66 | A | 7.97588222 | 0.1897 |
| 67 | B | 7.93881251 | 49.4827 |
| 68 | B | 7.87824296 | 24.0121 |
| 69 | A | 7.82390005 | 289.2797 |
| 70 | A | 7.71224685 | 62.6043 |
| 71 | B | 7.57513734 | 258.571 |
| 72 | B | 7.44940841 | 36.8803 |
| 73 | A | 7.4406304 | 0.1248 |
| 74 | B | 7.36462829 | 58.0227 |
| 75 | A | 7.26920712 | 102.9819 |
| 76 | B | 7.26689949 | 23.4804 |
| 77 | B | 7.24181688 | 83.7932 |
| 78 | A | 7.09016378 | 3.7945 |
| 79 | A | 7.02878645 | 15.203 |
| 80 | B | 6.98309313 | 2.1444 |
| 81 | B | 6.86499217 | 13.7586 |
| 82 | A | 6.84221730 | 15.1946 |
| 83 | A | 6.72288351 | 1.5501 |
| 84 | A | 6.62131864 | 19.7922 |
| 85 | B | 6.60255991 | 43.2698 |
| 86 | A | 6.4851474 | 60.5678 |
| 87 | B | 6.46454552 | 239.7199 |
| 88 | B | 3.25254924 | 0.4922 |
| 89 | A | 3.25190834 | 0.4489 |
| 90 | B | 3.25026212 | 2.2571 |
| 91 | A | 3.25020656 | 0.7123 |
| 92 | B | 3.23918786 | 2.7907 |
| 93 | A | 3.2382037 | 2.6755 |
| 94 | B | 3.23171955 | 7.6982 |
| 95 | A | 3.23169234 | 0.1246 |
| 96 | B | 3.21464771 | 12.7718 |
| 97 | A | 3.21389777 | 0.1229 |
| 98 | A | 3.19430118 | 42.5184 |
| 99 | B | 3.19386244 | 12.9836 |

## 6, VIBRATIONAL MODE ANALYSIS OF $C_{23}H_{12}^{++}$

Vibrational mode analysis using Gaussian09 package was summarized in Table 2, which was classified as follows,

(1) 3.19μm (calculated and corrected value): C-H stretching mode at two pentagon sites.
(2) 6.46μm: C-C stretching at all hexagon sites.
  Calculated wavelength was longer than observed one by 0.26μm.
(3) 8.97-9.19μm: Not identified in observed spectra
(4) 11.4-12.9μm: C-H out-of-plane bending and C-C stretching
(5) 13.5μm: C-H out-of-plane bending at pentagon site
(6) 14.3μm: C-H out-of-plane bending at pentagons and C-C stretching.

These vibration mode features are almost similar to common PAH mode analysis on a review paper (see page 1028, Tielen; 2013)

## 7, CONCLUSION

In order to analyze vibrational mode of void induced coronene molecules, density functional theory was applied. Among them, di-cation $C_{23}H_{12}^{++}$ is a possible carrier of the astronomically observed polycyclic aromatic hydrocarbon (PAH).

(1) Based on unrestricted density functional theory, multiple spin state analysis was done for neutral void coronene $C_{23}H_{12}$. Singlet spin state was most stable (lowest total energy).

(2) By the Jahn-Teller effect, there occurs serious molecular deformation. Point group D6h of pure coronene transformed to C2 symmetry in $C_{23}H_{12}$ having carbon two pentagons.

(3) Advanced singlet stable molecules were di-cation $C_{23}H_{12}^{++}$ and di-anion $C_{23}H_{12}^{--}$. Molecular configuration was almost similar with neutral $C_{23}H_{12}$.

(4) Electric dipole moment of $C_{23}H_{12}^{++}$ and $C_{23}H_{12}^{--}$ show reversed direction each other with 1.19 and 2.63 Debey.

(5) Calculated infrared harmonic wavelength and relative peak height of $C_{23}H_{12}^{++}$ show a very likeness to observed one of two astronomical sources of HD44179 and NGC7027. Observed emission peaks of 3.3, 6.2, 7.6, 7.8, 8.6, 11.2, 12.7, 13.5 and 14.3μm were compared well with single molecule $C_{23}H_{12}^{++}$ calculated peaks of 3.2, 6.5, 7.6, 7.8, 8.7, 11.4, 12.9, 13.5, and 14.3μm.

(6) Harmonic vibrational mode analysis was done for $C_{23}H_{12}^{++}$. At wavelength 3.2 μm, C-H stretching at pentagons was featured. From 6.4 to 8.7μm, C-C stretching mode was observed. In-plane-bending of C-H was in a range of 7.6-9.2μm. Both C-H out-of plane bending and C-C stretching were accompanied from 11.4 to 14.3μm.
.

## REFERENCES

Armus, L., Charmandaris, V., Bernard-Salas, J., et al. 2007, ApJ, 656, 148
Bauschlicher, C. W., & Langhoff, S. R. 1997, Spectrochim. Acta A, 53, 1225
Bauschlicher, C. W., Peeters, E., & Allamandola, L. J. 2008, ApJ, 678, 316
Bauschlicher, C. W., Peeters, E., & Allamandola, L. J. 2009, ApJ, 697,
Becke, A. D. 1993, J. Chem. Phys., 98, 5648
Boersma, C., Bregman, J.D. & Allamandola, L. J.. 2013, ApJ, 769, 117
Boersma, C., Bauschlicher, C. W., Ricca, A., et al. 2014, ApJ Supplement Series, 211:8
Boersma, C., Mattioda, A. L., Bauschlicher JR, C. W., et al. 2009, ApJ, 690, 1208
Cook, D. J., Schlemmer, S., Balucani, N., et al. 1998, J. Phys. Chem., 102, 1467
de Frees, D. J., Miller, M. D., Talbi, D., Pauzat, F., & Ellinger, Y. 1993, ApJ, 408, 530
Engelbracht, C. W., Kundurthy, P., Gordon, K. D., et al. 2006, ApJ, 642, L127
Frisch, M. J., Pople, J. A., & Binkley, J. S. 1984, J. Chem. Phys., 80, 3265
Frisch, M. J., Trucks, G.W., Schlegel, H. B., et al. 2009, Gaussian 09, Revision A.02 (Wallingford, CT: Gaussian, Inc.)
Geballe, T. R., Tielens, A. G. G. M., Allamandola, L. J., Moorhouse, A., & Brand, P. W. J. L. 1989, ApJ, 341, 278
Hudgins, D. M., & Allamandola, L. J. 1999a, ApJ, 516, L41
Hudgins, D. M., & Allamandola, L. J. 1999b, ApJ, 513, L69
Kim, H.-S., Wagner, D. R., & Saykally, R. J. 2001, Phys. Rev. Lett., 86, 5691
Langhoff, S. R. 1996, J. Phys. Chem., 100, 2819
Malloci, G., Joblin, C., & Mulas, G. 2007, Chem. Phys., 332, 353
Meeus, G., Waters, L. B. F. M., Bouwman, J., et al 2001, A&A, 365, 476
Moutou, C., Leger, A., & D'Hendecourt, L. 1996, A&A, 310, 297
Moutou, C., Sellgren, K., Verstraete, L., & L´eger, A. 1999, A&A, 347, 949
Ota, N., Gorjizadeh, N. & Kawazoe, Y. , 2011, Journal of Magnetics Society of Japan, 35, 414
Ota, N. 2014a, arXiv org., 1408.6061
Ota, N. 2014b, arXiv org., 1412.0009
Oomens, J. 2011, in In PAHs and the Universe: A Symposium to Celebrate the 25th Anniversary of the PAH Hypothesis, EAS Publications Series, ed. A. G.G. M. Tielens & C. Joblin (Cambridge: Cambridge University Press), 46, 61
Oomens, J., Sartakov, B. G., Tielens, A. G. G. M., Meijer, G., & von Helden, G. 2001, ApJ, 560, L99
Oomens, J., Tielens, A. G. G. M., Sartakov, B. G., von Helden, G., & Meijer, G. 2003, ApJ, 591, 968
Pathak, A., & Rastogi, S. 2007, Spectrochim. Acta A, 67, 898
Peeters, E., Hony, S., van Kerckhoven, C., et al. 2002, A&A, 390, 1089
Piest, H., von Helden, G., & Meijer, G. 1999, ApJ, 520, L75
Regan, M. W., Thornley, M. D., Bendo, G. J., et al. 2004, ApJS, 154, 204
Ricca, A., Bauschlicher, C. W., Jr., Boersma, C ., Tielens ,A. & Allamandola, L. J. 2012, ApJ, 754, 75
Ricca, A., Bauschlicher, C. W., Jr., Mattioda, A. L., Boersma, C., & Allamandola, L. J. 2010, ApJ, 709, 42
Ricca, A., Bauschlicher, C. W., Jr., & Allamandola, L. J. 2011b, ApJ, 729, 94
Schlemmer, S., Cook, D. J., Harrison, J. A., et al. 1994, Science, 265, 1686
Sellgren, K., Uchida, K. I., & Werner, M. W. 2007, ApJ, 659, 1338
Smith, J. D. T., Draine, B. T., Dale, D. A., et al. 2007, ApJ, 656, 770
Szczepanski, J., & Vala, M. 1993a, ApJ, 414, 646   Phys., 14, 2381
Tielens, A, G, G, M 2013, Rev. Mod. Phys., 85, 1021
Verstraete, L., Puget, J. L., Falgarone, E., et al. 1996, A&A, 315, L337